# Estimation of transmitted wavefronts at defocused positions in a broad bandwidth range

**Haoyu Wang[1], Qiyuan Zhang[2,3], Fang Wang[1], Chan Li[1], Yiming Zhu[1], Songlin Zhuang[1], Yanhui Kang[4], Yongjin Zhu[5], Dahai Yu[6], Sen Han[1,3*]**

[1] *School of Optical-Electrical and Computer Engineering, University of Shanghai for Science and Technology, Shanghai 200093, China*
[2] *School of Optoelectronic Engineering, Changchun University of Science and Technology, Changchun 130022, China*
[3] *Suzhou H&L Instruments LLC, Suzhou 215123, China*
[4] *National Institute of Metrology, Beijing 100029, China*
[5] *Shanghai Supore instruments Co Ltd., Shanghai 200444, China*
[6] *Focused Photonics Inc., Hangzhou 310052, China*
E-mail: senhanemail@126.com

**Abstract:** Although laser interferometers have emerged as the main tool for measurement of transmitted wavefronts, their application is greatly limited, as they are typically designed for operation at specific wavelengths. In a previous study, we proposed a method for determining the wavefront transmitted by an optical system at any wavelength in a certain band. Although this method works well for most monochromatic systems, where the image plane is at the focal point for the transmission wavelength, for general multi-color systems, it is more practical to measure the wavefront at the defocused image plane. Hence, in this paper, we have developed a complete method for determining transmitted wavefronts in a broad bandwidth at any defocused position, enabling wavefront measurements for multi-color systems. We also conducted experiments to verify these assumptions, validating the new method. The experimental setup has been improved so that it can handle multi-color systems, and a detailed experimental process is summarized. With this technique, application of broadband transmission wavefront measurement can be extended to most general optical systems, which is of great significance for characterization of achromatic and apochromatic optical lenses.

## 1. Introduction

    Laser interferometry has become the main method for transmitted wavefront measurement, due to its high accuracy [1–4]. However, laser interferometers are typically optimized and built for only one wavelength [5–9], to test different optical systems, whose operating bands vary from extreme ultraviolet to far-infrared, a number of laser interferometers with different design wavelengths are necessary, which largely increases the cost of this testing. In view of the above, we developed a method for estimating the transmitted wavefront at any wavelength in a certain band [10]. The technique requires the construction of a function relating the transmitted wavefront (expressed using Zernike polynomials [11, 12]) to wavelength, using reference wavefronts from three or four wavelengths. Two expressions were obtained that can be used to represent the relationship between Zernike coefficients and wavelength: the Conrady formula, and a new formula named the apochromatic characteristic formula (ACF). These formulas can describe the wavelength dependent Zernike coefficients at the focal point of an optical system. While the focal point is usually the best working position for monochromatic systems (such that it is typically defined as the image plane), in most cases, for multi-color systems such as achromatic systems, the image plane is not at the focal point of a given design wavelength [13–15]; there is a small shift in the focal point for constituent wavelengths in relation to the image plane. Therefore, for most general optical transmission systems, it is more practical to measure the wavefront at the image plane (defocused plane).

    In this paper, we propose methods for estimating the characteristics of transmitted wavefronts at defocused positions, for a broad bandwidth range. To do this, we analyzed the relationship between Zernike coefficients at the defocused plane at different wavelengths. At the same time, we demonstrate that in a short defocus range, the relationship between Zernike coefficients and position is approximately linear. Hence, these coefficients can be determined for a given defocused position, demonstrating that broadband characterization of transmitted wavefronts at defocused positions is possible. Our simulations reveal that this method is particularly effective for estimating the $Z_3$ coefficient, which relates to the defocus characteristic

of a wavefront, while other Zernike coefficients at defocused positions can be derived directly using the Conrady formula, such that the wavefront at that position can be reconstructed. Our demonstration further proves the accuracy of the Conrady formula and the ACF in describing the wavelength-dependent behavior of the main parameters of an optical system (such as Zernike coefficients and focal lengths). We also give a study of the capability of the Conrady and the ACF with defocus algorithms in different waveband, which has important significance in practical measurements. These analyses are then verified experimentally with study of an achromatic doublet's image plane. We have improved the experimental setup so that it can measure such multi-color systems, and a detailed experimental process is summarized.

This paper establishes and completes the method of estimating broadband wavefront for most general multi-color systems (including achromatic and apochromatic systems where the defocus plays an important role). While this method is important in practical wavefront measurements, it can also help the process in the optical design. As the wavefront contains a lot of information of the optical system such as the geometrical aberration, modulation transfer function (MTF) and point spread function, the wavelength dependent Zernike coefficients developed in this paper can perform faster analysis for these properties in a broad waveband than using ray tracing. More specifically, in the traditional optical design, the MTF is only calculated at one or several specific wavelengths. With the wavelength dependent Zernike coefficients, it is easy and fast to calculate wavefronts at a large number of wavelengths, which can give us an understanding of an overall behavior of the transmission system in a broad bandwidth.

## 2. Principle of operation

A schematic illustration of a multi-wavelength transmission system is shown in Fig. 1. Here, the back focal lengths of beams with wavelengths defined as $\lambda_1$, $\lambda_2$, and $\lambda_3$ are $l_1$, $l_2$, and $l_3$, respectively. Mathematical models for describing Zernike coefficients at the focal position as functions of wavelength, specifically, the Conrady formula and the ACF, were derived in our previous research, and are written as,

$$Z_i(\lambda) = A_i + \frac{B_i}{\lambda} + \frac{C_i}{\lambda^{3.5}}, \tag{1}$$

$$Z_i(\lambda) = A_i + \frac{B_i}{\lambda^{2.8}} + \frac{C_i}{\lambda^{5.8}} + D_i \lambda^{0.3}, \tag{2}$$

where $Z_i(\lambda)$ is the wavefront's $i$th order Zernike coefficient at the focal position, for the given wavelength, $\lambda$. In the above, (1) is the Conrady formula, while (2) is the ACF. The inspiration of the Conrady formula and the ACF came from dispersion formulas, based on the fact that the change of the transmitted wavefront Zernike coefficients with the wavelength is essentially caused by the change of the refractive index. The Conrady formula is suitable for modeling most monochromatic systems and some achromatic systems where the behavior of Zernike coefficients has one inflection point at most. The ACF can be used for more complex systems including apochromatic systems, for curves with up to two inflection points. More information about the basic principle of the Conrady formula and the ACF can be found in [10]. Since the ACF is a more general formula, the power exponents for the wavelength terms can be tuned according to the optical system, to improve prediction accuracy. As the ACF requires data at one more wavelength than the Conrady formula, in some fast estimations where a very broad bandwidth is not necessary, the latter is a suitable alternative. Using these equations, for a wavelength $\lambda_n$, the wavefront at the corresponding focal position, $l_n$, can be predicted. However, in practice, for most cases, the wavefronts for all wavelengths at a given position, for example, at the image plane of the system, $l_0$, are required. With some exceptions dictated by different types of achromatism, this fixed position does not coincide with any wavelength's focal point. Therefore, a method for estimating wavefronts at defocused positions is required. The method developed, termed the defocus algorithm, is described below.

As the magnitude of defocus in the design waveband is usually small, we assumed a linear relationship between Zernike coefficients and positions in the defocus range. Thus, the Zernike coefficient at defocused points can be estimated from the following expression:

$$Z_i(\lambda, \Delta d) = Z_i(\lambda) + k_i(\lambda) \cdot \Delta d(\lambda), \tag{3}$$

where $Z_i(\lambda)$ and $\lambda$, are as defined in (1) and (2), $\Delta d(\lambda)$ is the defocus at the given wavelength, and $k_i(\lambda)$ is the change in $Z_i(\lambda)$ per unit length at the given wavelength. As we have assumed a linear relationship between Zernike coefficients and position, it can be surmised that $k_i(\lambda)$ and $Z_i(\lambda)$ have similar behaviors. In the following sections, we describe the results of simulations and experiments conducted to validate the assumptions made in defining this technique.

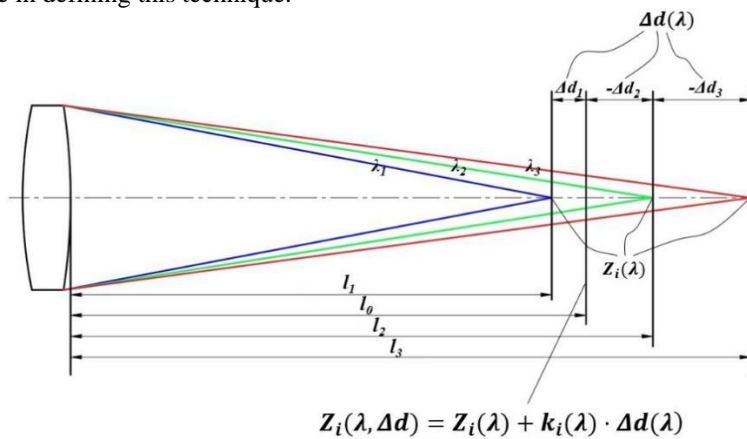

$$Z_i(\lambda, \Delta d) = Z_i(\lambda) + k_i(\lambda) \cdot \Delta d(\lambda)$$

Fig. 1. Schematic of a multi-wavelength transmission system. Beams with different wavelengths ($\lambda_1$, $\lambda_2$, and $\lambda_3$) are focused at different positions ($l_1$, $l_2$, and $l_3$), while the image plane of the system is at $l_0$. The amount of defocus from $l_0$ for wavelength $\lambda_n$ is described as $\Delta d_n$. In a practical optical system, the value of $\Delta d_n$ in the design waveband is usually small (from tens of micrometers to several millimeters depending on the depth of focus).

## 3. Simulation

To verify the operation of the defocus algorithm, we conducted numerical simulations on an achromatic doublet lens using the Zemax software package. For these simulations, we chose a typical achromatic doublet telephoto objective with an *f*-number of *f*/8, constructed from K9 and ZF1 glass, from the lens catalog (Lens Design 4th Edition). Design wavelengths of 480 nm, 510 nm, 546.1 nm, 590 nm, and 643.8 nm were selected, with weights of 0.3, 0.6, 1, 0.6, and 0.3, respectively. The fields of view (FOVs) were set to 0 °, 0.4 ° and 0.75°. Following optimization, the system's back focal length (i.e., the location of the image plane) was 1196.162 mm.

### 3.1 Direct predictions using the Conrady formula

To assess the suitability of the Conrady formula for predicting Zernike coefficients at defocused positions, we measured the first nine of these in Zemax, for wavefronts at both the focal and the image planes. For systems more complex than the doublet lens (particularly when there are multiple inflection points in the Zernike coefficient-wavelength curve), the ACF should be used for extrapolation. In these simulations, we conducted measurements at 10 nm intervals in the 400 nm–1000 nm range, for all FOVs. These nine coefficients correspond to the primary aberrations summarized in Table 1. Following this, to solve the Conrady formula, we selected datapoints at 550 nm, 600 nm, and 650 nm, enabling prediction of Zernike coefficients at arbitrary wavelengths within the 400 nm–1000 nm range. A comparison of the results of Zemax simulations with predictions made using the Conrady formula is shown in Fig. 2. A summary of which coefficients the Conrady formula is capable of predicting is also given in Table 1.

**Table 1. Summary of the Conrady formula's ability to predict Zernike coefficients at defocused positions**

| FOV | $Z_0$ | $Z_1$ | $Z_2$ | $Z_3$ | $Z_4$ | $Z_5$ | $Z_6$ | $Z_7$ | $Z_8$ |
|---|---|---|---|---|---|---|---|---|---|
| | piston | tilt | | defocus | astigmatism | | coma | | spherical |
| 0° | - | 0 | 0 | × | 0 | 0 | 0 | 0 | √ |
| Non-zero | - | √ | √ | × | √ | √ | √ | √ | √ |

Coefficients marked with "√" can be estimated by the Conrady formula directly; those marked with "×" cannot. Prediction of Zernike coefficients corresponding to tilt, astigmatism, and coma for the 0° FOV is relatively trivial; as these aberrations are zero at this FOV, the related coefficients should also be zero.

Fig. 2 depicts the wavelength-dependent nature of the $Z_3$, $Z_7$, and $Z_8$ coefficients transmitted at a 0.75 ° FOV, for a representative comparison between the results of Zemax simulation (solid lines) and Conrady formula prediction (dashed lines). In these plots, blue lines represent Zernike coefficients at each wavelength's respective focal point, while red lines represent coefficients at the fixed image plane. The datapoints selected for solution of the Conrady formula are marked as circles. Here, we note that for all FOVs, the $Z_3$ term is non-zero, regardless of whether it was measured at the focal point. This is because of a slight defocus applied in consideration of image quality optimization. From this image, it can be seen that the Conrady formula can be used to determine a wavefront's Zernike coefficients at the focal point corresponding to the transmission wavelength. The technique is similarly successful for 0 ° and other small angle FOVs. However, at the image plane, where wavefronts from most wavelengths are defocused, the Conrady formula fails to predict some Zernike coefficients accurately. This failure was noted for the $Z_0$ coefficient, which corresponds to piston aberration (not shown in Fig. 2). Since this term is usually removed in wavefront measurement, this particular failure is trivial. However, the Conrady formula produces inaccurate predictions of the $Z_3$ coefficient (corresponding to defocus) at the image plane, which requires addressing. As the remaining Zernike coefficients representing the primary aberrations of an optical system ($Z_1$, $Z_2$, $Z_4$, $Z_5$, $Z_6$, $Z_7$ and $Z_8$) are relatively independent of defocus, they can be estimated by the Conrady formula directly. However, since the values of coefficients measured at defocused positions differ from those measured at the focal position (see Fig. 2(b)), the Conrady formula is only accurate if the initial references are measured at the defocused position required.

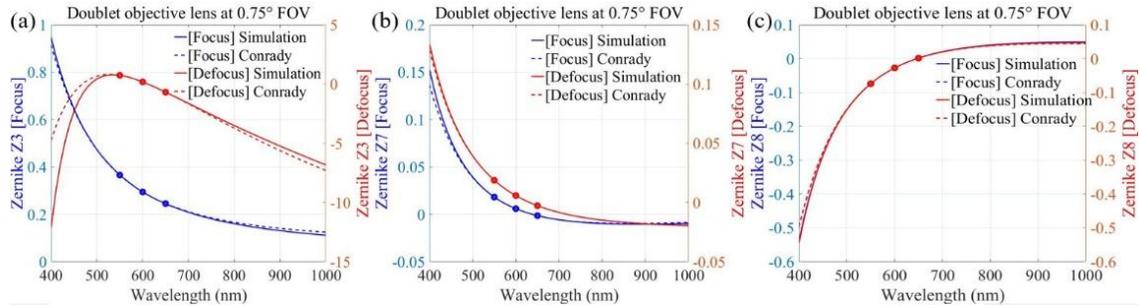

Fig. 2. Prediction of (a) $Z_3$, (b) $Z_7$, and (c) $Z_8$ coefficients using the Conrady formula for a wavefront transmitted at the 0.75° FOV. Curves in blue represent results at the focal plane, while curves in red represent results at the defocused image plane.

*3.2 Dependence of Zernike coefficients on defocus*

The simulation discussed above indicated that $Z_0$ and $Z_3$ have a strong dependence on defocus. Since the imaging quality of most optical systems is acceptable in the range of four times the depth of focus [16], calculation of wavefronts beyond it is of little significance. Hence, using Zemax, we acquired $Z_3$ coefficients in the range of four times the depth of focus in 0.5 mm intervals from each wavelength's focal point. Results for wavefronts transmitted at 500 nm, 550 nm, 600 nm, and 650 nm are shown in Fig. 3(a). The image demonstrates that a predominantly linear change could be observed in $Z_3$ with respect to distance, within the small defocus range. Therefore, recalling Equation (3), $k_i(\lambda)$ and $Z_i(\lambda)$ should have similar dependences on wavelength, meaning that the Conrady formula can be used to describe the behavior of $k_i(\lambda)$

$$k_i(\lambda) = A_{ki} + \frac{B_{ki}}{\lambda} + \frac{C_{ki}}{\lambda^{3.5}} . \qquad (4)$$

To verify this assumption, we compared the simulated values of $k_3(\lambda)$ to those predicted using a solved Conrady formula (based on datapoints collected at 550 nm, 600 nm, and 650 nm), observing good alignment between the two, as shown in Fig. 3(b). Although a similar linear dependence on distance was noted for the $Z_4$–$Z_8$ coefficients, this data is not pictured, as the change over the defocus range was negligible.

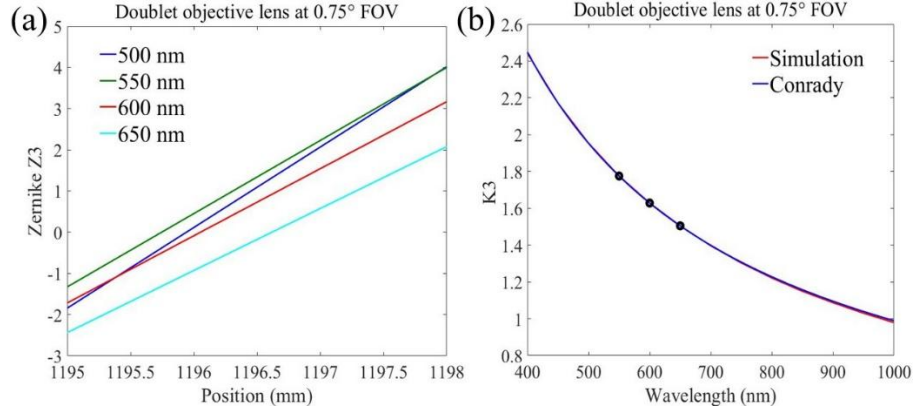

Fig. 3. (a) Behavior of $Z_3$ as a function of position. A linear trend is observed for each wavelength considered. (b) Comparison of the $k_3$-wavelength relationship suggested by Zemax and the Conrady formula. The curves are practically identical when the Conrady formula is derived from datapoints at 550 nm, 600 nm, and 650 nm.

### 3.3 Wavelength-dependent defocus

The defocus for each wavelength is dependent on the position of the image plane of the system, $l_0$, and the back focal length (which, like the focal length, can be expressed as a function of wavelength [17-18] as $l(\lambda)$), and is written as,

$$\Delta d(\lambda) = l_0 - l(\lambda). \tag{5}$$

The optical design principle indicates that, in theory, the image plane of a system can be determined by a combination of its design wavelengths and their weights, using the following equation,

$$l_0 = \frac{\sum_1^n w_n l_n}{\sum_1^n w_n}, \tag{6}$$

where $l_n$ is the back focal length at a given design wavelength, and $w_n$ is the weight for that wavelength. Although determination of the image plane of modern optical systems is usually very complex (requiring a comprehensive consideration of their purpose, and aberrations in the optical system), in the simulation under discussion, the position of the image plane is calculated using Equation (6), since the FOVs considered are small enough to ensure that their effect is much smaller than that of the change in wavelength.

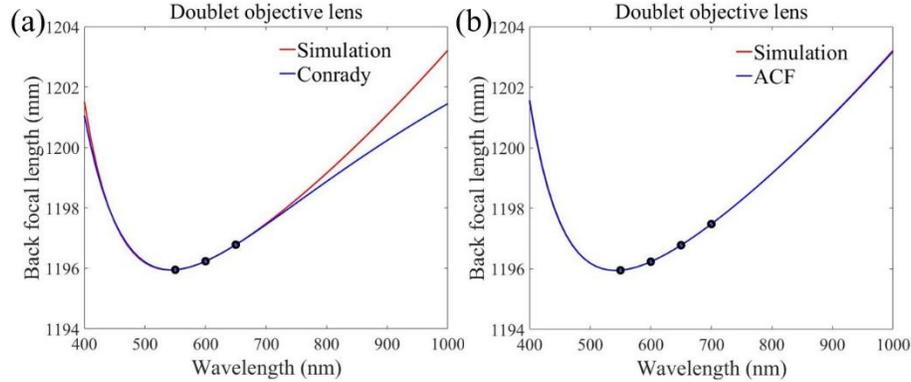

Fig. 4. Wavelength-dependent back focal length estimated using (a) the Conrady formula, and (b) the apochromatic characteristic formula (ACF).

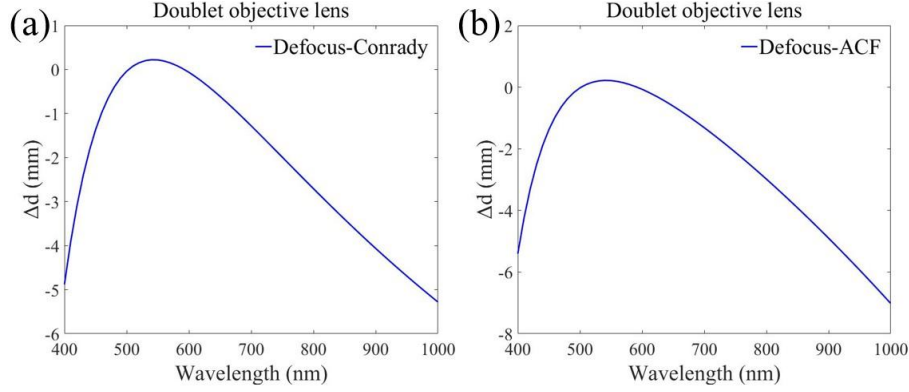

Fig. 5. Wavelength-dependent defocus calculated using (a) the defocus-Conrady formula, and (b) the defocus-ACF. In these figures, the position of the image plane was calculated using Equation (6), and the defocus was subsequently derived according to Equation (5). For Fig. 5(a), $\Delta d(\lambda)$ was calculated by solving the Conrady formula at 550 nm, 600 nm, and 650 nm. For Fig. 5(b), $\Delta d(\lambda)$ was calculated by solving the ACF at 550 nm, 600 nm, 650 nm and 700 nm.

Most achromatic systems have focal shift curves with similar characteristics, i.e., a curve with a single inflection point. Since, from our previous study, the Conrady formula is able to model the behavior of such curves, we used this to provide an initial estimate of the relationship between back focal length and wavelength, as shown in Fig. 4(a). As before, the curve was derived using datapoints at 550 nm, 600 nm, and 650 nm. It can be seen that the solution starts to deviate from the simulation from around 700 nm. Hence, we derived a model using the ACF (shown in Fig. 4(b)), since it performs broadband estimation much better. Following the adjustment of the power exponents that is characteristic of achromatic systems, the following formula for back focal length was obtained:

$$l(\lambda) = A_l + \frac{B_l}{\lambda^{2.9}} + \frac{C_l}{\lambda^{5.9}} + D_l \lambda^{1.9}, \tag{7}$$

with datapoints at 550 nm, 600 nm, 650 nm, and 700 nm selected for solution of the above equation.

Once functions relating back focal length to wavelength were derived with both the Conrady formula and the ACF, we used these to calculate back focal lengths at the design wavelengths (480 nm, 510 nm, 546.1 nm, 590 nm and 643.8 nm). Using Equation (6), the positions of the image plane were subsequently derived as 1196.170 mm, using the results from the Conrady formula, and 1196.164 mm (ACF), both of which are very close to the Zemax optimized value (1196.162 mm). Finally, the defocus at each wavelength was derived according to Equation (5), as shown in Fig. 5.

### 3.4 Applying the defocus algorithm

Using the defocus algorithm given in Equation (3), we estimated the Zernike coefficient for wavefronts at the image plane, using the same datapoints as in Sections 3.2 and 3.3. The results of these predictions for the $Z_3$ coefficient are shown in Fig. 6. A comparison with Fig. 2(a) shows that the defocus algorithm produces results with much better accuracy than predictions made using the Conrady formula directly. The defocus-ACF shows good performance throughout the 400 nm to 1000 nm waveband. While the defocus-Conrady formula's accuracy is limited to a narrower band, it requires less data for prediction.

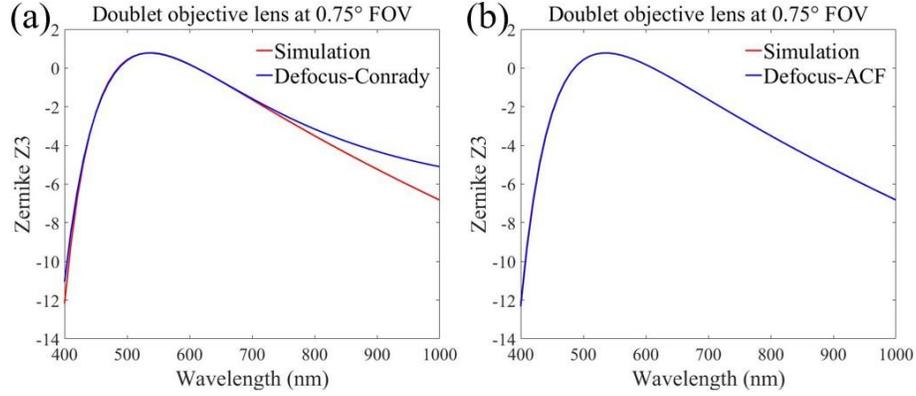

Fig. 6. $Z_3$ coefficients predicted according to the defocus algorithm using (a) the Conrady formula, and (b) the ACF.

Table 2. Maximum residual error for prediction of the first nine Zernike coefficients at defocused positions, for a 0.75 °FOV

|  | Conrady | Defocus-Conrady | Defocus-ACF |
|---|---|---|---|
| $Z_0$ | -7.37 | -1.73 | 0.20 |
| $Z_1$ | / | / | / |
| $Z_2$ | 0.010 | 0.029 | 0.032 |
| $Z_3$ | -7.42 | -1.73 | 0.15 |
| $Z_4$ | -0.0025 | -0.0015 | -0.0013 |
| $Z_5$ | / | / | / |
| $Z_6$ | / | / | / |
| $Z_7$ | 0.0054 | 0.015 | 0.016 |
| $Z_8$ | -0.049 | -0.045 | -0.045 |

Table 2 shows a comparison of the three different prediction methods (direct solution of the Conrady formula for Zernike coefficients at the image plane (Method A), the defocus-Conrady formula (Method B), the defocus-ACF (Method C)) with respect to the maximum residual error generated for each Zernike coefficient. While Method A is suitable for estimating the $Z_4$–$Z_8$ coefficients, with Methods B and C, the accuracy of the $Z_3$ is significantly improved. To demonstrate this, we plotted the residual errors for prediction of the $Z_3$ coefficient in the waveband from 400 nm to 1000 nm using the different methods in Fig. 7. The image illustrates that the defocus-ACF is accurate in the whole waveband. In contrast, while the Conrady formula is accurate for wavelengths above 500 nm, it produces large errors for wavelengths between 400 nm and 500 nm. Similarly, the defocus-Conrady formula produces relatively accurate results at short wavelengths (~500 nm–700 nm) but generates large errors in the band above 700 nm due to errors in back focal length estimation in this range (see Fig. 4(a)).

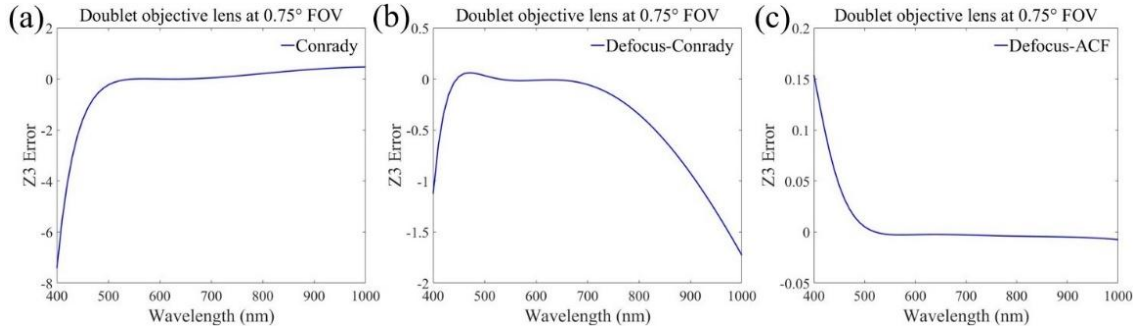

Fig. 7. Residual error in defocused $Z_3$ estimates predicted using (a) the Conrady formula, (b) the defocus-Conrady formula, and (c) the defocus-ACF.

## 4. Experimental procedures

In brief, we performed interferometric measurements of the achromatic doublet at five different wavelengths, to obtain Zernike coefficients describing wavefronts at both the focal plane and at defocused positions. Using similar extrapolation techniques to those described in our previous research, Zernike coefficients for wavefronts transmitted at wavelengths differing from these references can be obtained.

### 4.1 Experimental setup

Fig. 8 shows the experimental setup used for measuring the wavefronts transmitted by the achromatic doublet lens. Here, the main components are a Fizeau phase-shifting interferometer with five changeable laser sources (532 nm, 561 nm, 632.8 nm, 671 nm, and 721 nm) and a high-quality concave spherical mirror on a translation stage. The achromatic doublet lens is placed behind the collimated beam, and the rotation of the spherical mirror around the $x$-axis and $y$-axis is tuned to minimize the values of the $Z_1$ and $Z_2$ terms, to make the $x$-tilt and $y$-tilt as close to zero as possible. Since, in this study, we have to measure Zernike coefficients of wavefronts at the focal point of each wavelength and several defocused positions, the longitudinal position of the high-quality spherical mirror should be able to be changed. It is also essential to make sure that the optical axis is well aligned to the rail axis before taking measurements (see Fig. 8(b) and 8(c)). Previously, we changed the longitudinal position of the spherical mirror using the 5-axis mount. However, we observed that the $z$-axis control provided by this component introduced large alignment errors, which changed $Z_1$ and $Z_2$ noticeably during movement. This is because the screws squeezing the mirror are on the edge of the mount, causing them to move inconsistently. To avoid this situation, we fastened the 5-axis mount to an extra one-axis translation stage (shown in Fig. 8(a) and Fig. 8(d)), to move the entire 5-axis mount back and forward during the measurement. This reduces the alignment error to an acceptable range.

### 4.2 Measurement procedure

In contrast to simulation, in experiments, we restricted measurement of transmitted wavefronts to the 0 ° FOV, where the $Z_4$–$Z_7$ terms should all equal zero. In spite of our alignment of the doublet and the spherical mirror (where $Z_1$ and $Z_2$ are both zero), this is not the case in practice, due to lens manufacture error (especially decentering error and tilt error) and residual setup alignment error. Therefore, to recover the wavefront, $Z_4$–$Z_8$ should be measured. A summary of the procedure for measurement of these terms is shown in Fig. 9.

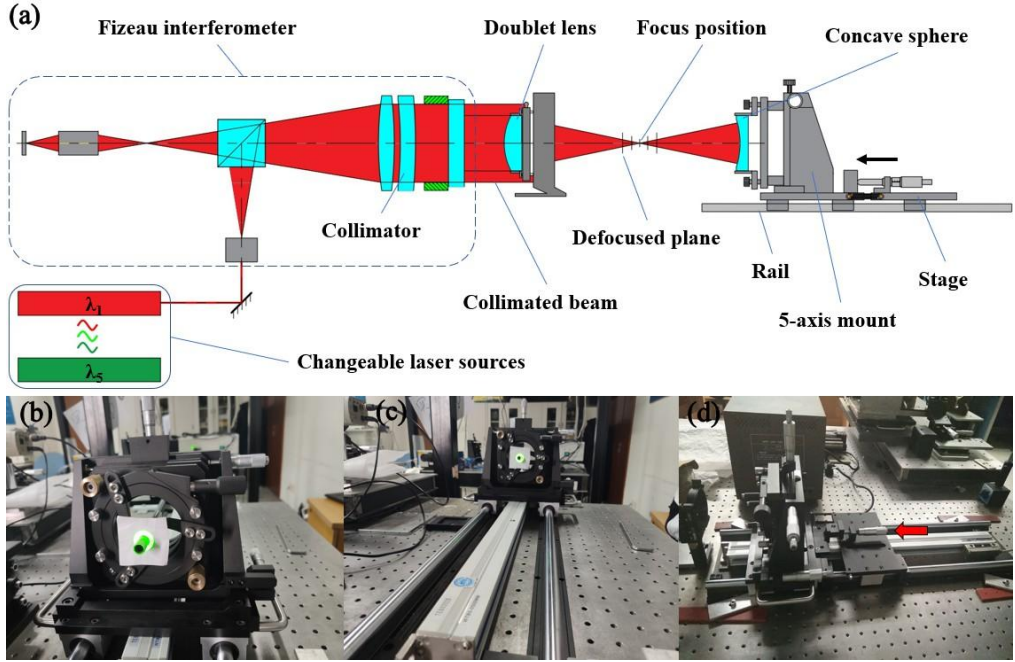

Fig. 8. (a) Schematic of the experimental setup used for measuring wavefronts at the focal point and defocused positions. (b) and (c) Illustration of the alignment process with beam spots. (d) One-axis translation stage used to improve control of the spherical mirror's longitudinal motion.

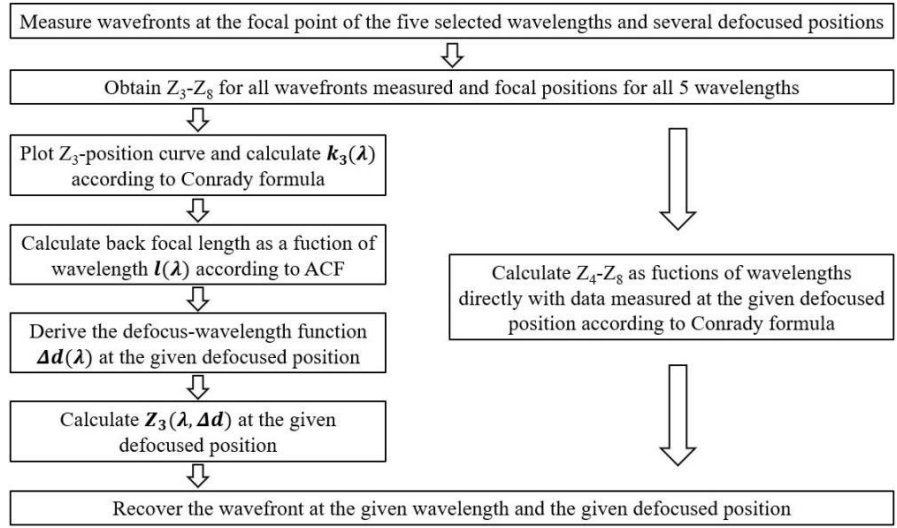

Fig. 9. Summary of measurement procedure for wavefront reconstruction.

## 5. Results and discussion

In this section, we discuss experiments conducted to explore the validity of the defocus algorithm, and assess potential problems associated with practical measurement.

*5.1 Wavefront measurements*

As directed by Fig. 9, we measured the $Z_4$ to $Z_8$ coefficients of wavefronts transmitted at 532 nm, 561 nm, 632.8 nm, 671 nm, and 721 nm, at a fixed image plane, and the respective focal planes for these wavelengths. For prediction of Zernike coefficients at arbitrary wavelengths, we investigated both curve fitting and solution of the Conrady formula, in consideration of possible measurement errors. In these experiments, datapoints from three of the five wavelengths were used to predict the Zernike coefficients at the other two wavelengths. The accuracy of both methods was subsequently characterized based on the proximity of these predictions to measurement. From previous experience, we observed that datapoints to be used in extrapolation should cover as large a range of the bandwidth as possible, to avoid large errors in broadband estimation due to errors in measurement. Hence, we used data from 532 nm, 632.8 nm, and 721 nm to estimate Zernike coefficients at 561 nm and 671 nm.

The results of these experiments are shown in Fig. 10. In this image, black circles indicate measurements, red lines indicate predictions made by solving the Conrady formula, and blue lines indicate predictions made by curve fitting. Here, we again see that the Conrady formula accurately describes the wavelength-dependent behavior of Zernike coefficients, based on the proximity of the predictions to measurements. In contrast to a conclusion made in our previous paper [10], we note here that the solution method produces better results than the fitting method. This discrepancy is because of the differences in the nature of most monochromatic systems (which was the focus of the previous study) and achromatic systems. For the former, fitting techniques are usually more accurate than solving techniques when there are errors in the measurement data, because Zernike coefficients have a predominantly linear relationship with wavelength in the measurement range; while both techniques produce similar results, the errors produced by solving the Conrady formula are more significant than those produced by curve fitting. With achromatic systems, the relationship of the Zernike coefficients with wavelength is more non-linear (e.g., $Z_6$ and $Z_8$ shown in Fig. 10(c) and Fig. 10(e)), resulting in larger differences between fitting and solving predictions. For such curves, the solution method produces results that are more consistent with the experimental data. We also note that there is a discrepancy in $Z_7$ between the measurement and the prediction. We think this is due to experimental errors in the measurement. As the change in $Z_7$ at different wavelength is very small, the measured result is more likely to be affected by the experimental error.

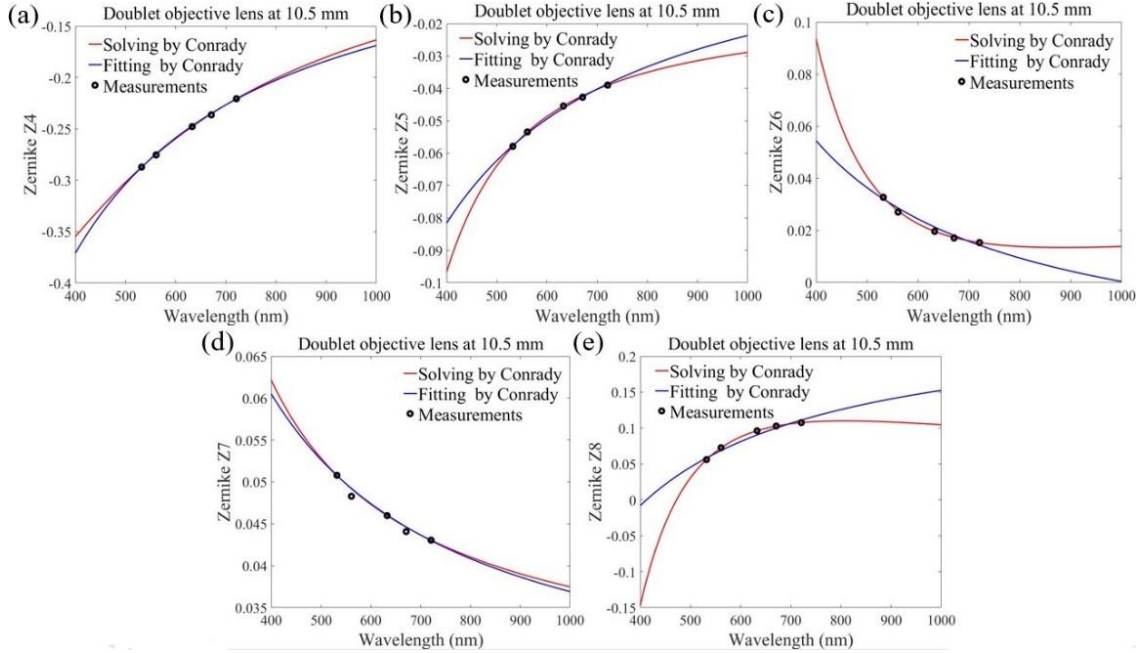

Fig. 10. Comparison of Zernike coefficient measurements with predictions obtained by solving the Conrady formula and fitting the Conrady formula for (a) $Z_4$, (b) $Z_5$, (c) $Z_6$, (d) $Z_7$, and (e) $Z_8$ terms. Curves were constructed from extrapolation of measurements at 532 nm, 632.8 nm, and 721 nm.

### 5.2 Measurement of $k_i(\lambda)$ and defocus

Our simulations indicate that of the eight Zernike coefficients typically used in wavelength measurement ($Z_0$ is usually ignored), only $Z_3$ has a strong dependence on defocus. Hence, only the $k_3$ parameter requires measurement. To do this, we measured the variation of $Z_3$ with position for all five laser wavelengths by changing the location of the concave mirror in the $z$-axis. The results of this measurement are as shown in Fig. 11(a). We noted significant errors in the value of $Z_3$ for measurements performed at positions close to a wavelength's focal point, because of the coupling between tilt and longitudinal motion. Hence, every time we modified the spherical mirror's longitudinal position, the $x$-tilt and $y$-tilt were also adjusted, to ensure $Z_1$ and $Z_2$ were always equal to zero. While the change in defocus accompanying this change in tilt is small in absolute terms, at positions close to the focal point, the relative change is large since the defocus is close to zero in these locations. Therefore, to derive the linear relationship between $Z_3$ and position, we considered only data obtained at locations where the defocus is much larger than zero. The $k_3$ parameter is given by the gradient of the resulting curve (Fig. 11(a)). As in Section 5.1, we used data obtained at 532 nm, 632.8 nm, and 721 nm to predict the values of $k_3$ at 561 nm and 671 nm. A comparison between results obtained by solving the Conrady formula and those obtained through curve fitting are shown in Fig. 11(b). Here, since the relationship between $k_3$ and wavelength is more linear (based on the results of simulation shown in Fig. 3(b)), curve fitting the Conrady formula produces more accurate predictions.

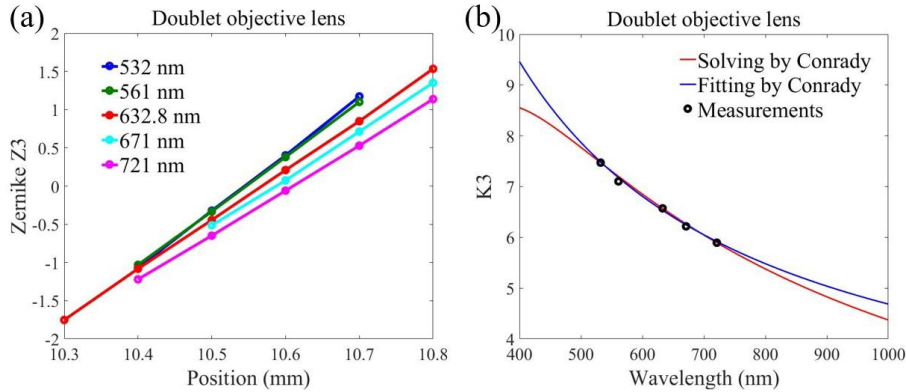

Fig. 11. (a) Variation of $Z_3$ with position for wavefronts transmitted at different wavelengths. The gradient of each line represents the value of $k_3$ for each wavelength. (b) Comparison of $k_3(\lambda)$ predicted by solving and fitting the Conrady formula. Curves were obtained from extrapolation of data at 532 nm, 632.8 nm and 721 nm.

The location of the focal point for the different wavelengths is required for full description of the defocus parameter. Typically, this location is defined as the position where $Z_3$ equals zero. However, to avoid the large errors obtained in measuring $Z_3$, caused by the spherical mirror's coupled tilt and longitudinal motion as discussed above, we extrapolated the focal point for each wavelength through a linear fit of the $Z_3$-position curves shown in Fig. 11(a). Table 3 shows a comparison of each wavelength's focal point obtained from direct measurement and from calculation. The maximum error between both sets of values is ~5 μm, which has a big influence in estimating $l(\lambda)$.

Table 3. Comparison of focal points obtained from measurement and through curve fitting for each laser wavelength

|  | 532 nm | 561 nm | 632.8 nm | 671 nm | 721 nm |
| --- | --- | --- | --- | --- | --- |
| Measurement (mm) | 10.548 | 10.545 | 10.569 | 10.582 | 10.607 |
| Linear fitting (mm) | 10.543 | 10.545 | 10.567 | 10.583 | 10.607 |

Our simulations indicated that the ACF is more effective than the Conrady formula in describing the wavelength dependent behavior of focal length and defocus. Hence, an additional datapoint is required for accurate estimation of these parameters. A comparison of the effect of the selected wavelength on prediction accuracy is shown in Fig. 12, where we consider three different models where the 561 nm, the 632.8 nm, and the 671 nm datapoints are omitted. Fig. 12(a) shows that the ACF's prediction of back focal length as a function of wavelength matches the simulation shown in Fig. 4(b), with small deviations appearing at short wavelengths and long wavelengths due to measurement errors. Fig. 12(b), which depicts the predicted defocus at an arbitrary position (10.5 mm), shows similar behavior. These results indicate that omitting the 632.8 nm (i.e., selecting datapoints at 532 nm, 561 nm, 671 nm, and 721 nm) or the 671 nm datapoint (i.e., selecting datapoints at 532 nm, 561 nm, 632.8 nm, and 721 nm) produces better results. Hence, we adopt the configuration omitting the 671 nm (in order to compare with the Conrady formula) datapoint in further experiments.

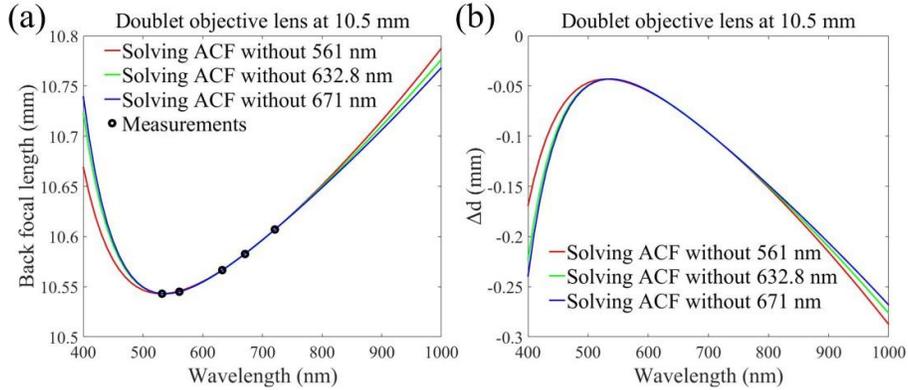

Fig. 12. (a) Effect of datapoint selection on wavelength-dependent back focal length predicted using the ACF. (b) Wavelength-dependent defocus corresponding to 12(a).

*5.3 Wavefront reconstruction*

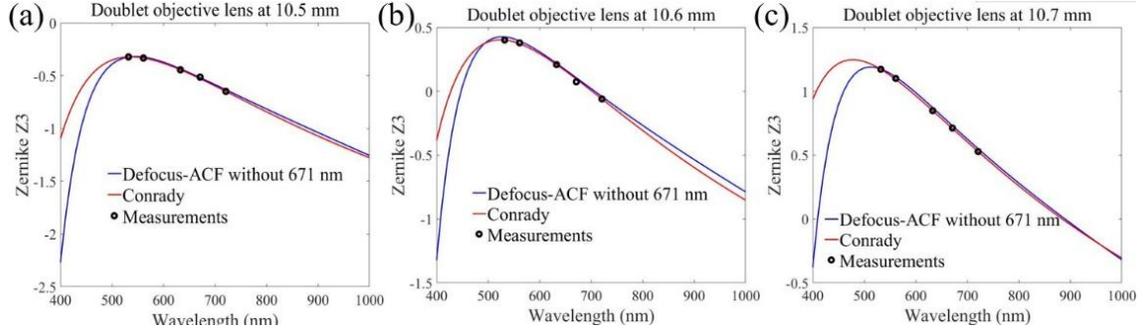

Fig. 13. Comparison of $Z_3$ coefficients at (a) 10.5 mm, (b) 10.6 mm, and (c) 10.7 mm predicted using the defocus-ACF (blue lines) and the Conrady formula (red lines). For the Conrady formula, reference measurements were obtained at 532 nm, 632.8 nm, and 721 nm, while an additional datapoint was collected at 561 nm for the defocus-ACF.

For further validation of the defocus algorithm (Equation (3)), we predicted the $Z_3$ coefficient at three arbitrary positions (10.5 mm, 10.6 mm and 10.7 mm) using the defocus-ACF method. Here, we selected datapoints at 532 nm, 632.8 nm, and 721 nm for modeling $k_3(\lambda)$ using the Conrady formula, while an additional datapoint was selected at 561 nm, for modeling $\Delta d(\lambda)$ according to the ACF. The results of these calculations are shown as blue lines in Fig. 13 (measurements are indicated as black circles). For comparison, we also present results given by direct estimation with the Conrady formula (Method A) in red lines, with datapoints at 532 nm, 632.8 nm and 721 nm used for extrapolation. We note that the defocus-ACF's prediction of the datapoint at 671 nm matches measurements, as an indication of its accuracy. For wavelengths above 500 nm, the Conrady formula is consistent with the defocus-ACF, which agrees with the simulation presented in Fig. 7. At short wavelengths (between 400 nm and 500 nm), the estimation provided by the defocus-ACF is more reliable.

A comparison of actual wavefronts measured at the three defocused positions (transmitted at 671 nm), with the wavefronts reconstructed using the $Z_3$–$Z_8$ estimates provided by Methods A and C is shown in Fig. 14, with residual errors summarized in Table 4. As there is little difference in their performance for wavelengths above 500 nm, both methods produce similar results at 671 nm. The image indicates that the predicted wavefronts match the measurement obtained at 10.5 mm and 10.7 mm. However, there is a deviation between the estimation and the measurement at 10.6 mm. We suggest that this is because, here, the estimated value of $Z_3$ is close to zero at 671 nm (see Fig. 13(b)), such that a small deviation from measurement creates a significant change in the wavefront. This result highlights the difficulty in estimating wavefronts at positions close to the transmission wavelength's focal point.

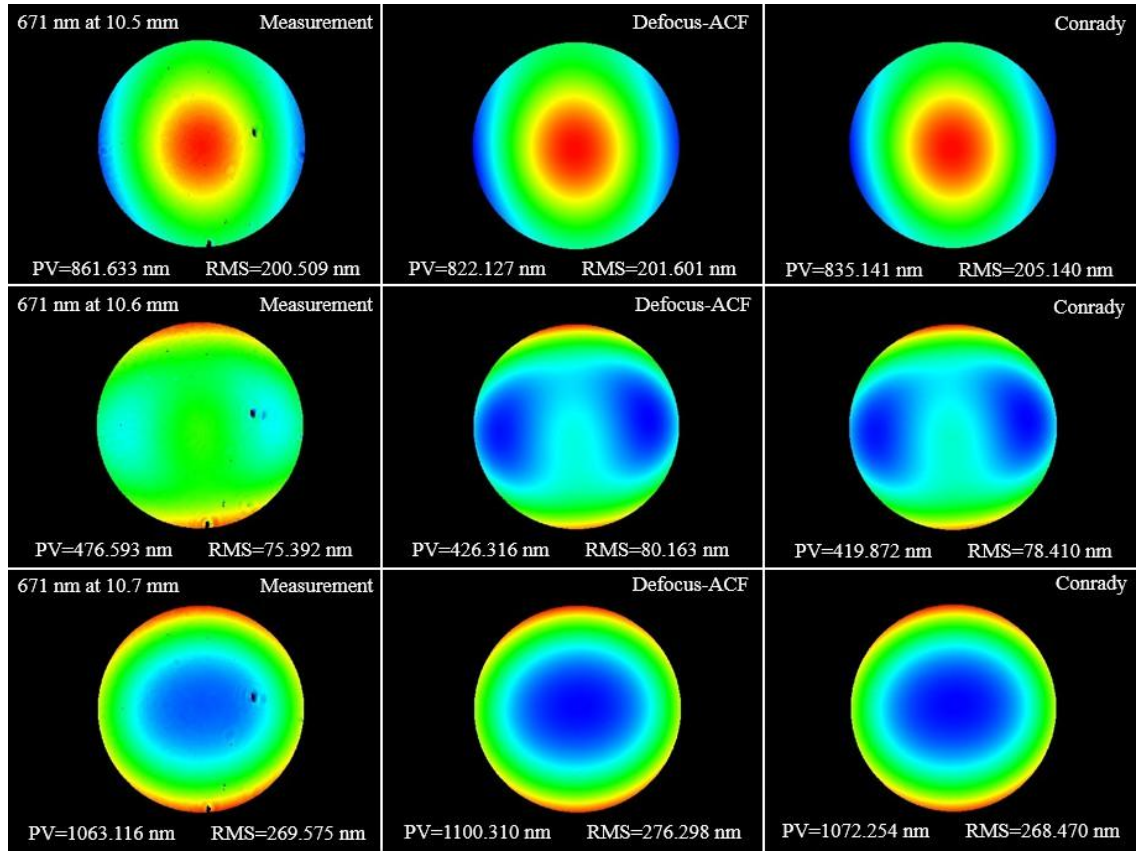

Fig. 14. Comparison between measured wavefronts and wavefronts reconstructed using Zernike coefficients estimated by the defocus-ACF and the Conrady formula.

**Table 4. Residual errors between predicted and reconstructed wavefronts**

|  | Defocus-ACF | | Conrady | |
|---|---|---|---|---|
| 10.5 mm | MAX=171.467 nm | RMS=10.055 nm | MAX=166.232 nm | RMS=11.295 nm |
| 10.6 mm | MAX=218.806 nm | RMS=15.217 nm | MAX=217.318 nm | RMS=12.498 nm |
| 10.7 mm | MAX=191.860 nm | RMS=12.436 nm | MAX=198.433 nm | RMS=10.494 nm |

Here, MAX refers to the maximum deviation, while RMS refers to the root mean square error.

## 6. Conclusion and discussion

In summary, in this paper we studied two different techniques for estimating defocused transmitted wavefronts in a broad bandwidth. With the first, Zernike coefficients at defocused positions were directly estimated with the Conrady formula, while the second, the defocus algorithm, requires the establishment of relationships between Zernike coefficients and wavelengths, as well as position. We detailed a defocus-Conrady formula and a defocus-ACF, named according to the model used for describing wavelength-dependent behavior. Simulations and experiments show that Zernike coefficients have a linear relationship with position in a small defocus range, a property which we used in determination of the focal position for a given transmission wavelength. Based on this, we demonstrated that Zernike coefficients at defocused points can be derived from coefficients at the focal point, further proving the accuracy of the Conrady formula and the apochromatic characteristic formula in describing the behavior of the main parameters of an optical system (such as Zernike coefficients and focal lengths). We also observed that $Z_3$ plays an important role in wavefront reconstruction, highlighted by the fact that the defocus algorithm is most effective for this coefficient.

Considering the application of this technique in practical situations, it is important to choose modes of operation that make measurements faster and easier. For example, most achromatic systems working in the visible waveband do not have a very broad bandwidth. Hence, the inaccurate performance of direct estimation of defocused Zernike coefficients using the Conrady formula for wavelengths below 500 nm is inherently avoided, and this technique is recommended. For accurate estimation at wavelengths below 500 nm, the defocus algorithm must be used, with the defocus-ACF proving more reliable throughout the 400 nm to 1000 nm range. For apochromatic systems, there are no choices but to use the defocus-ACF with measurements at four wavelengths. Another point worth pointing out is that we strongly recommend using the ACF when dealing with systems with small F numbers, as the Conrady formula gives larger errors at wavelengths in the edge of 400 nm to 1000 nm waveband. Note that it may be not sufficient to use only the first 9 terms of the Zernike coefficients. Higher order terms should be considered to describe the accurate wavefront aberration. Since real measurements feature unavoidable errors, we also conducted preliminary studies to analyze the difference between solving and fitting the Conrady formula, which will be expanded in the future. The method proposed in this paper is of great significance for evaluating the qualities of multi-color systems.

## Funding

This work was partially supported by the National Natural Science Foundation of China (Grants No. 11804224), National Key Research and Development Program of China (Grants No. 2016YFF0101903).

## Acknowledgments

The authors thank Jianjun Hou for helping with setting up the experimental device. The authors also thank Xueyuan Li for writing the software to obtain wavefronts and Zernike coefficients from the interferometer.

## References

1. W. Zhu, L. Chen, Y. Yang, R. Zhang, D. Zheng, Z. Han, and J. Li, "Advanced simultaneous phase-shifting Fizeau interferometer," Opt. Lasers. Tech. 111, 134–139 (2019).
2. D. Zheng, L. Chen, J. Li, Q. Sun, W. Zhu, J. Anderson, J. Zhao, and A. Schülzgen, "Circular carrier squeezing interferometry: Suppressing phase shift error in simultaneous phase-shifting point-diffraction interferometer," Opt. Lasers. Eng. 102, 136–142 (2018).
3. Y. Zong, J. Li, M. Duan, G. Chen, W. Lu, R. Zhu, and L. Chen, "Dynamic phase-deforming interferometry: suppression of errors from vibration and air turbulence," Opt. Lett. 44(16), 3960–3963 (2019).
4. J. S. Lee, H. S. Yang, and J. W. Hahn, "Wavefront error measurement of high-numerical-aperture optics with a Shack–Hartmann sensor and a point source," Appl. Opt. 46(9), 1411–1415 (2007).
5. P. L. Domenicali, "Infrared laser interferometer," Proc. SPIE 0192, 6–12 (1979).
6. D. G. Flagello, and B. Geh, "Lithographic lens testing: analysis of measured aerial images, interferometric data, and photoresist measurements," Proc. SPIE 2726, 1–11 (1996).
7. Y. Zhu, K. Sugisaki, M. Okada, K. Otaki, Z. Liu, J. Kawakami, M. Ishii, J. Saito, K. Murakami, M. Hasegawa, C. Ouchi, S. Kato, T. Hasegawa, A. Suzuki, H. Yokota, and M. Niibe, "Wavefront measurement interferometry at the operational wavelength of extreme-ultraviolet lithography," Appl. Opt. 46(27), 6783–6792 (2007).
8. S. Xu, B. Tao, Y. Guo, and G. Li, "Polarization aberration analysis in lithographic tools," Opt. Eng. 58(8), 082405 (2019).
9. W. Peng, C. Ho, W. Lin, Z. Yu, C. Huang, C. Kuo, and W. Hsu, "Design, tolerance analysis, fabrication, and testing of a 6-in. dual-wavelength transmission sphere for a Fizeau interferometer," Opt. Eng. 56(3), 035105 (2017).
10. Q. Zhang, H. Wang, P. Wu, Y. Fu, X. Li, Q. Wang, and S. Han, "Estimating transmitted wavefronts in a broad bandwidth based on Zernike coefficients," J. Opt. 21(9), 095601 (2019).
11. K. Liu, J. Wang, H. Wang, and Y. Li, "Wavefront reconstruction for multi-lateral shearing interferometry using difference Zernike polynomials fitting," Opt. Lasers. Eng. 106, 75–81 (2018).
12. V. V. Orlov, V. Y. Venediktov, A. V. Gorelaya, E.V. Shubenkova, and D. Z. Zhamalatdinov, "Measurement of Zernike mode amplitude by the wavefront sensor, based on the Fourier-hologram of the diffuse scattered mode," Opt. Lasers. Tech. 116, 214–218 (2019).
13. S. E. Ivanov, and G. E. Romanova, "Optical material selection method for an apochromatic athermalized optical system," J. Opt. Tech. 83(12), 729–733 (2016).
14. T, Lim, and S. Park, "Achromatic and athermal lens design by redistributing the element powers on an athermal glass map," Opt. Express 24(16), 18049–18057 (2016).
15. J. L. Rayces, M. Rosete-Aguilar, "Selection of glasses for achromatic doublets with reduced secondary spectrum. I. Tolerance conditions for secondary spectrum, spherochromatism, and fifth-order spherical aberration," Appl. Opt. 40(31), 5663–5676 (2001).
16. W. J. Smith, "Modern Optical Engineering, 4th ed," (Mcgraw-Hill, 2008).
17. K. Seong, and J. E. Greivenkamp, "Chromatic aberration measurement for transmission interferometric testing," Appl. Opt. 47(35), 6508–6511 (2008).
18. X. Jiang, J. A. Kuchenbecker, P. Touch, and R. Sabesan, "Measuring and compensating for ocular longitudinal chromatic aberration," Optica 6(8), 981–990 (2019).